%% file: main.tex
\documentclass[10pt, conference]{IEEEtran}
\IEEEoverridecommandlockouts

\usepackage{cite}
\usepackage{amsmath,amssymb,amsfonts}
\usepackage{algorithmic}
\usepackage{graphicx}
\usepackage{subcaption}
\usepackage{textcomp}
\usepackage{color}
\usepackage{booktabs}
\usepackage[normalem]{ulem}
\usepackage{enumitem}
\usepackage{multirow}
\usepackage{url}
\usepackage{hyperref}
\usepackage[table,xcdraw]{xcolor} 

\definecolor{brightturquoise}{rgb}{0.85, 1, 1}

\def\BibTeX{{\rm B\kern-.05em{\sc i\kern-.025em b}\kern-.08em
    T\kern-.1667em\lower.7ex\hbox{E}\kern-.125emX}}
\begin{document}

\title{Time and Tokens: Benchmarking End-to-End Speech Dysfluency Detection\\
}

\author{
\IEEEauthorblockN{\textit{Xuanru Zhou$^{1}$, Jiachen Lian$^{2}$, Cheol Jun Cho$^{2}$, Jingwen Liu$^{1}$, Zongli Ye$^{1}$, Jinming Zhang$^{1}$, Brittany Morin$^{3}$} \\ 
\textit{David Baquirin$^{3}$, Jet Vonk$^{3}$, Zoe Ezzes$^{3}$ \ Zachary Miller$^{3}$, Maria Luisa Gorno-Tempini$^{3}$, Gopala Anumanchipalli$^{2\dagger}$}}
\\
$^{1}$Zhejiang University \  $^{2}$UC Berkeley \ $^{3}$UCSF\\
\small \tt xuanruzhou15@gmail.com,  \{jiachenlian, gopala\}@berkeley.edu
}


\maketitle

\input{abstract}
\input{intro}
\input{yolo}

\input{token}

\input{experiment}
\input{conclusion}

\bibliographystyle{IEEEtran}
\bibliography{refs}

\end{document}

%% file: abstract.tex
\begin{abstract}
Speech dysfluency modeling is a task to detect dysfluencies in speech, such as repetition, block, insertion, replacement, and deletion. Most recent advancements [1-3] treat this problem as a \textit{time-based object detection} problem. In this work, we revisit this problem from a new perspective: tokenizing dysfluencies and modeling the detection problem as a \textit{token-based} automatic speech recognition (ASR) problem. We propose rule-based speech and text dysfluency simulators and develop \textit{VCTK-token}, and then develop a Whisper-like seq2seq architecture to build a new benchmark with decent performance. We also systematically compare our proposed token-based methods with time-based methods, and propose a unified benchmark to facilitate future research endeavors. We open-source these resources for the broader scientific community. The project page is available at \url{https://rorizzz.github.io/}.
 
\end{abstract}

\begin{IEEEkeywords}
dysfluency, time, token, simulation, detection, end-to-end, clinical
\end{IEEEkeywords}

%% file: intro.tex
\section{Introduction}
Recent developments in automatic speech recognition (ASR)\cite{radford2023whisper, zhang2023google-usm, pratap2023scaling-speech} have achieved human parity in rich-resource languages, however, \textit{only at the word level}. Human speech is a hierarchical form of words, phonemes, prosody, dysfluencies, etc., and research in understanding those non-word cues, which are essential parts of human conversations, is still limited. 
This work focuses on a specific non-word cue: \textit{dysfluency}\footnote{dysfluency is interchangable with disfluency}. Dysfluencies are defined as a set of non-fluencies such as sound repetition, insertion, deletion, replacement, or hesitation\cite{lian2023unconstrained-udm}. Understanding dysfluencies in speech can help understand the intent of speech in dialogue systems and also has a huge impact on speech therapy and speech disorder screening~\cite{lian2023unconstrained-udm}.

Early work on dysfluency modeling primarily focused on stutter detection, which was expressed as a speech classification problem~\cite{CHIAAI20122157, LPCC, ESMAILI2016104, 10068490, Detect-lstm, fluentnet2021, alharbi2020segment-detection3, segment-detection4, DBLP:conf/interspeech/BayerlWNR22, howell1995automatic,10094692}. However, there are several important limitations. First, stutter is just one aspect of dysfluencies. Second, there can be multiple dysfluencies in one utterance. Third, dysfluencies are text-dependent~\cite{lian2023unconstrained-udm}. Fourth, accurately locating timing information of dysfluencies is essential in clinical applications~\cite{lian2023unconstrained-udm}. Fifth, the datasets used by these methods are of small scale, and the annotations are still subject to the aforementioned limitations.
To overcome these limitations, UDM~\cite{lian2023unconstrained-udm} proposed modeling the dysfluency problem as a \textit{time-based} object detection problem: identifying what the dysfluencies are and where they occur. H-UDM~\cite{lian2024hierarchical} boosted UDM via a dynamic word boundary updating algorithm. YOLO-Stutter~\cite{zhou2024yolo} directly adapts YOLO~\cite{redmon2016look} to simulated speech-text alignments\cite{kim2021conditional-vits} and detects the dysfluencies in an end-to-end manner. 

\begin{figure}[ht]
    \centering
    \includegraphics[height=5.5cm]{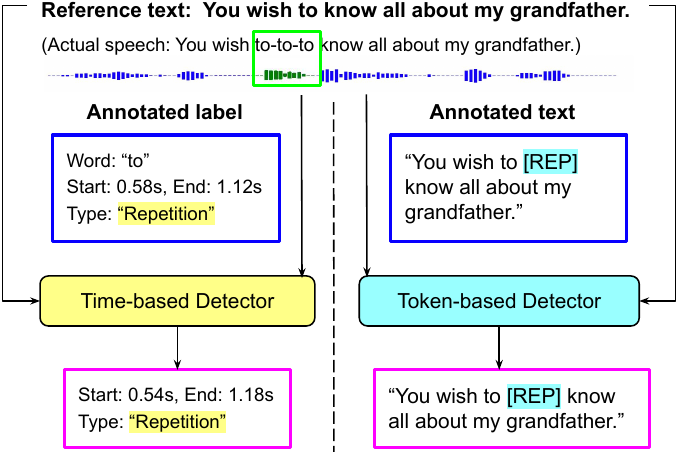}
    \caption{Comparison of Time-based and Token-based Methods}
    \label{method-comparison}
\end{figure}

\begin{figure*}[htb]
\begin{minipage}[b]{0.7\linewidth}
  \centering
\centerline{\includegraphics[width=13.5cm]{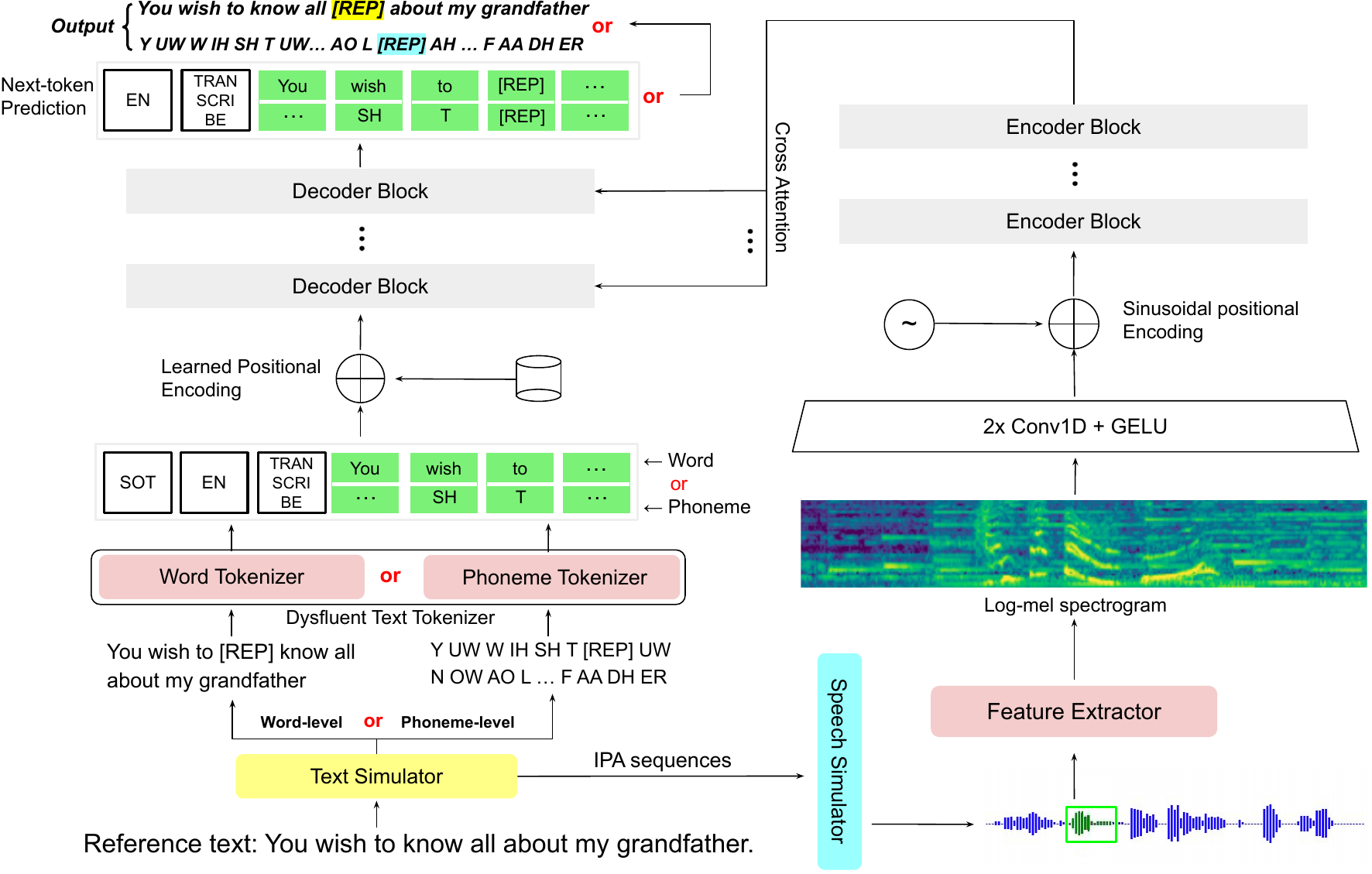}}
\vspace{5pt}
  \centerline{(a) Pipeline of the Token-based Dysfluency Detection}\medskip
\end{minipage}
\hfill
\begin{minipage}[b]{0.2\linewidth}
\centering
\centerline{\includegraphics[width=4.1cm]{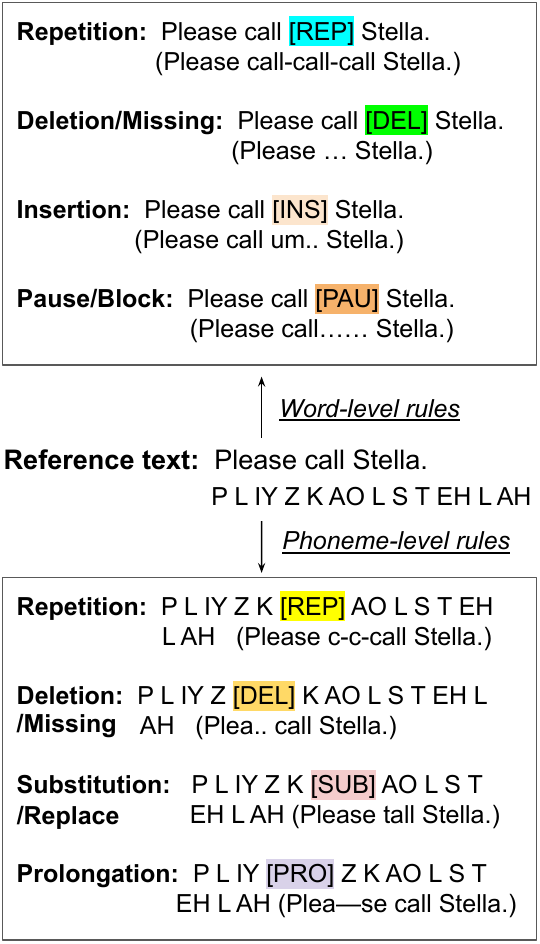}}
\vspace{23pt}
  \centerline{(b) Text Simulator }\medskip
\end{minipage}
\caption{We begin by using a text simulator to convert the reference text into annotated text, either at the word \textbf{or} phoneme level. Next, we generate corresponding dysfluent speech through a speech simulator. The Whisper feature extractor processes the resulting audio waveform, while the Whisper Tokenizer \textbf{or} CMU Phoneme Tokenizer handles the word or phoneme-level annotated text. These audio and text representations are then fed into a Whisper-like encoder-decoder architecture for training and prediction. \textbf{a)} shows the entire pipeline for token-based dysfluency detection. \textbf{b)} illustrates the rules for injecting tokenizing dysfluency into the text space. Here are audio samples of dysfluent speech 
 \url{https://bit.ly/3XI5CTu}
}
\label{fig-2}
\end{figure*}

While \textit{time-based} methods exhibit huge potential, we approach this problem from a fresh perspective: \textit{token-based} modeling. We list our intuitions as follows. First, token sequences are more straightforward to understand than time-based predictions, as visualized in Fig.~\ref{method-comparison}. Second, all types of dysfluencies are tokenizable, and both type and time information can still be derived from token sequences. 
To design this \textit{token-based} pipeline, we follow Yolo-Stutter~\cite{zhou2024yolo} for the simulation pipeline and use the same dysfluency rules (Text Simulator) to generate dysfluent text at both word and phoneme levels. Then, a speech simulator is employed to generate dysfluent speech samples, with the ground truth annotations being simply the dysfluent text (word or phoneme). We name this new dataset \textit{VCTK-Token}.
In the subsequent phase, we implement the Whisper paradigm~\cite{radford2022whisper}, which processes dysfluent speech as input and generates predictions for dysfluent text tokens (comprising both word or phoneme sequences and dysfluency markers), as illustrated in Fig. \ref{fig-2}. Both time-aware and token-aware metrics are developed to form a new benchmark with decent performance. 

Lastly, we aim to unify both \textit{time-based} and \textit{token-based} methods. To do so, we first propose the integration of longest common subsequence (LCS) prior to enhance the \textit{time-based} method~\cite{zhou2024yolo}, which we call \textit{YOLO-Stutter-LCS}. Then, we create a comprehensive collection of simulated datasets (VCTK-Stutter~\cite{zhou2024yolo}, VCTK-TTS~\cite{zhou2024yolo}, and VCTK-Token). We also develop a joint \textit{time-based} and \textit{token-based} metric for unified evaluation. Results on both simulated data and disordered speech suggest that \textit{token-based} methods outperform \textit{time-based} methods in most metrics. To facilitate research in end-to-end dysfluency modeling and provide a platform for benchmark purposes, we open-source our work at \url{https://rorizzz.github.io/}.

%% file: yolo.tex
\section{Time-based Dysfluency Detection}
Current de facto methods for dysfluency modeling are \textit{time-based} methods~\cite{lian2023unconstrained-udm, lian2024hierarchical, zhou2024yolo}: detecting the type of dysfluency in speech and returning its time range. We will review the state-of-the-art end-to-end method YOLO-Stutter~\cite{zhou2024yolo} and also propose \textit{YOLO-Stutter-LCS}, which
includes the longest common subsequence (LCS) algorithm as alignment constraint to boost dysfluency detection. 

\subsection{YOLO-Stutter}
YOLO-Stutter~\cite{zhou2024yolo} takes speech-text alignment~\cite{kim2021conditional-vits} as input, followed by a spatial encoder and a temporal encoder to predict dysfluency regions through a 1D adaptation of the YOLO objective~\cite{redmon2016look}. For each timestep, YOLO-Stutter predicts boundary scores, dysfluency class scores, and dysfluency confidence scores. To handle the data scarcity issue, it simulates dysfluent speech in the acoustic domain (VCTK-Stutter) and text domain (VCTK-TTS) respectively. We take YOLO-Stutter as our \textit{time-based} baseline and further develop an enhanced version, \textit{YOLO-Stutter-LCS}, by incorporating additional LCS dysfluent alignment.

\subsection{YOLO-Stutter-LCS}
Dysfluencies usually have aligned targets. For example, when the ground truth text is "Please call stella" and the speech is "Please c-c-call s-stalla", to detect dysfluency, we need to examine the pronunciation of "call" and "stella" respectively. We found that by applying the longest common subsequence (LCS)~\cite{wiki-lcs} algorithm on two sequences, the dysfluencies automatically align to their corresponding words (e.g., c-c-call → call, s-stalla → stella). This is due to its local sequence alignment property: the cost function is only decided by a subsequence (the first alignment), in contrast to global sequence aligners such as DTW. Based on this intuition, we pre-segment the data used in YOLO-Stutter via the LCS algorithm so that speech-text alignment can capture more dysfluency-related information. We name this method \textit{YOLO-Stutter-LCS}.

%% file: token.tex
\section{Token-based Dysfluency Detection}

\subsection{Dysfluency Simulation}

\subsubsection{Text Simulator}

We inject the tokenizing dysfluencies into text space at both the word and phoneme levels, obtaining annotated texts and their corresponding IPA sequences for generating dysfluent speech. The rules for injecting word or phoneme-level dysfluencies and deriving IPA sequences are consistent with \cite{zhou2024yolo}. Notably, in this work, we also introduce 6 dysfluency tokens, as illustrated in Fig. \ref{fig-2}(b), to denote the presence, type, and position of dysfluency within the text.





\subsubsection{Speech Simulator}
Similar to the method pipeline described in \cite{zhou2024yolo}, we employ the VITS \cite{kim2021conditional-vits} as our speech synthesizer. We feed IPA sequences generated by the text simulator into the VITS model to produce dysfluent speech. These \textit{(dysfluent speech, word / phoneme-level annotated text)} pairs serve as training data for our token-based detector.

\subsection{Speech Dysfluency Detector}
We treat the task of \textit{token-based} dysfluency detection as an automatic speech recognition (ASR) problem. 
We adopt Whisper architecture~\cite{radford2022whisper} which accurately predicts output tokens, including reference speech transcription and the dysfluency tokens at both word and phoneme levels.
To perform the detection, we require several essential components including \textit{Feature Extractor} and \textit{Dysfluent Text Tokenizer} to process dysfluent speech and annotated text respectively. Finally, a 
\textit{Whisper Detector} is applied for dysfluency prediction. Details are discussed below, and the overall pipeline is illustrated in Fig. \ref{fig-2}(a).

\subsubsection{Feature Extractor}
We employ the default Whisper~\cite{radford2022whisper} feature extractor, which outputs a log-mel spectrogram using an n\_fft of 400 and a hop\_length of 160 on 16 kHz waveform.

\subsubsection{Dysfluent Text Tokenizer}
We employ word-level tokenizer and phoneme-level tokenizer on dysfluent text from \textit{Text Simulator} respectively, as shown in Fig.~\ref{fig-2}(b).

\begin{itemize}[leftmargin=*]
\item \textbf{Word Tokenizer:} 
We utilize the default Whisper tokenizer~\cite{radford2022whisper} which takes annotated text as input and outputs tokens that correspond to the indices of the predicted text within its predefined vocabulary dictionary. The vocabulary comprises 50,258 vocabulary tokens along with additional special tokens to designate language, tasks, etc. Additionally, we incorporate 4 dysfluency tokens: \textit{[REP], [DEL], [INS]}, and \textit{[PAU]} into the vocabulary to enable its addressing of dysflunecy information.

\item  \textbf{Phoneme Tokenizer:}
Since the phoneme-level annotated text is in CMU phoneme~\cite{cmu-phoneme-dict} format, we employ the CMU phoneme tokenizer. Its vocabulary consists of 39 stress-free CMU phonemes, added by four dysfluency tokens: \textit{[REP], [DEL], [SUB]}, and \textit{[PRO]}. Additionally, we retain the special tokens from the default Whisper tokenizer.

\end{itemize}

\subsubsection{Whisper Detector}
We employ the Whisper-small~\cite{radford2022whisper} seq2seq architecture. It maps a sequence of audio spectrogram features to a sequence of dysfluency-injected text tokens. The encoder processes speech features to generate a series of hidden representations. The decoder autoregressively generates text tokens based on both hidden states from the encoder and previously generated tokens. We also modify the input dimension of the embedding layer and the output dimension of the last projection layer in the decoder to align with the updated or custom dictionary size (word or phoneme).

%% file: experiment.tex
\vspace{-5pt}

\begin{table}[h]
    \caption{Statistics and MOS of Simulated Datasets}
    \vspace{-3pt}
    \label{dysfluency-stats-vctk-tts}
    \centering
    \setlength{\tabcolsep}{6pt} 
    \renewcommand{\arraystretch}{1.1} 
    \resizebox{8.5cm}{!}{
    \begin{tabular}{l| c c c c} 
    \toprule
    Dysfluency & \textit{VCTK-Stutter}~\cite{zhou2024yolo} & 
    \textit{VCTK-TTS}~\cite{zhou2024yolo} & \textit{VCTK-Token}  \\
    \hline
    \hline
    Repetition & 121.80 & 114.92 & 146.09 \\
    Missing/Deletion & 61.05  &  122.05 & 146.04 \\
    Block/Pause & 61.05 & 68.56  & 75.47 \\
    Replace/Substitution & -- & 64.33 & 70.47  \\
    Prolongation & 60.75  & 57.06 & 70.43  \\
    Insertion & -- & -- & 75.55 \\
    \hline
    Total hours& 304.66  & 426.93  & 584.05\\
    MOS & 2.22 ± 0.84 & 4.13 ± 0.56 & 4.07 ± 0.43\\
    \bottomrule
    \end{tabular}}
    \label{statistics}
\end{table}

\vspace{-5pt}
\section{Experiments} \label{sec:experiments}
\vspace{-5pt}
\subsection{Datasets}
\textbf{1) VCTK-Token} 
is a VITS-based\cite{zhou2024yolo} simulated dataset \textit{proposed in this work}, extended from the VCTK~\cite{yamagishi2019cstr-vctk}. It comprises pairs of dysfluent speech and annotated text.
\textbf{2) VCTK-Stutter \cite{zhou2024yolo}} is a rule-based simulated dataset extended from VCTK, where dysfluency is directly injected into the extended acoustic space.
\textbf{3) VCTK-TTS \cite{zhou2024yolo}} is a VITS-based simulated dataset extended from VCTK. Dysfluencies are injected into the text space, and a text-to-speech model is used to generate the dysfluent speech.
\textbf{4) Aphasia Speech~\cite{zhou2024yolo}} We apply real dysfluent speech data comprises 38 English speakers diagnosed with Primary Progressive Aphasia (PPA).

\textit{VCTK-Stutter} and \textit{VCTK-TTS} are simulated datasets for time-based dysfluency detection. In order to evaluate the rationality and naturalness, we collected the Mean Opinion Score (MOS, 1-5) ratings from 10 people for three simulated datasets. Statistics for these datasets and MOS are presented in Table.\ref{statistics}.
Notably, since both \textit{VCTK-TTS} and \textit{VCTK-Token} speech samples are generated via VITS-based method\cite{zhou2024yolo}and extended by VCTK~\cite{yamagishi2019cstr-vctk}, they consistently exhibit similar content and dysfluency characteristics. This similarity is substantiated by the comparable MOS ratings.

\begin{table*}[h]
    \caption{Comparison of Time and Token-based Methods: Evaluation on Simulated Datasets and Aphasia Speech}
    \label{evaluate-comparison}
    \centering
    \setlength{\tabcolsep}{7pt} 
    \renewcommand{\arraystretch}{1.3} 
    \resizebox{17cm}{!}{
    \small
    \begin{tabular}{l l|c c| c c| c c |c c |c c| c c} 
     \toprule
      Methods& Evaluated Dataset & \multicolumn{2}{c|}{Repetition}& \multicolumn{2}{c|}{Missing / Deletion} & \multicolumn{2}{c|}{Block / Pause} & \multicolumn{2}{c|}{Replace / Substitution} & \multicolumn{2}{c|}{Prolongation} & \multicolumn{2}{c}{Insertion}\\
      \rowcolor{brightturquoise}
    & & EAcc.\% & CAcc.\% & EAcc.\% & CAcc.\% & EAcc.\% & CAcc.\% & EAcc.\% & CAcc.\% & EAcc.\% & CAcc.\% & EAcc.\% & CAcc.\% \\
    \hline
    \hline
    Time-based~\cite{zhou2024yolo} & VCTK-TTS& 90.07 & 98.78 & 86.13 &70.00 & 89.29 & \textbf{98.71} & 85.97 & 73.33 & 90.66& 93.74 & -- &-- \\
    Token-based & VCTK-Token& \textbf{99.76} & \textbf{99.29} & \textbf{98.41} & \textbf{97.53} & \textbf{99.04} & 98.28 & \textbf{96.11} & \textbf{95.75} & \textbf{98.89} &\textbf{98.81} & 98.93 & 97.47 \\
    \midrule
    Time-based~\cite{zhou2024yolo}&Aphasia Speech & 87.46 & \textbf{79.35} & 87.22 & 45.45 & 88.35 & \textbf{92.54} & 86.52 & 25.00 & 88.27 & \textbf{81.93} & -- &-- \\
    Token-based& Aphasia Speech& \textbf{93.02} & 72.73 & \textbf{92.61} & \textbf{81.82} & \textbf{92.34} & 54.74 & \textbf{94.21} & \textbf{86.67} & \textbf{90.86} &  44.44 & 92.33 & 82.35\\
    \bottomrule
    \end{tabular}}
\end{table*}

\vspace{-5pt}
\subsection{Metrics}
\textbf{1) Token Error Rate (TER)} measures the accuracy of transcribed text by comparing it to a reference text. It calculates the percentage of errors in terms of substitutions, deletions, and insertions.
\textbf{2) Dysfluency Exist Accuracy (EAcc.)} measures the accuracy of detecting whether dysfluencies are present in speech utterances.
It is defined by the proportion of correctly identified instances in all evaluated utterances. 
\textbf{3) Dysfluency Class Accuracy (CAcc.)} measures the accuracy of correctly detecting types of dysfluencies in utterances. It is defined by the proportion of accurately predicted dysfluent instances in all evaluated utterances.
\textbf{4) Bound Loss (BL)\cite{zhou2024yolo}} is the mean squared loss between the predicted and the actual bounds of the dysfluent regions on a time scale using a 20ms sampling frequency.
\textbf{5) Token Distance(TD)} measures the token-level displacement between predicted and actual dysfluency positions in text.

\subsection{Training Details}
The Whisper detector was trained using the CTC-Loss criterion and AdamW optimizer, incorporating gradient norm clipping and a linear learning rate decay to zero after a first 500 warmup steps, with a inital learning rate is 1e-5.
The data was split into a training and testing ratio of 90/10 and processed in batches of 8 per device.
For word-level dysfluency detection, we finetuned the pretrained Whisper-small checkpoint for 6,000 steps. For phoneme-level dysfluency detection, we trained the Whisper detector from scratch for 200,000 steps. 
All models were trained on 8 RTX A6000 GPUs, taking approximately 18 hours for word-level and 67 hours for phoneme-level training.

\subsection{Time-based Detection and YOLO-Stutter-LCS}
We trained the Time-based dysfluency detector~\cite{zhou2024yolo} on pre-segment VCTK-TTS datasets using LCS algorithm. To assess performance improvements, we evaluated YOLO-Stutter and YOLO-Stutter-LCS on two simulated datasets used in \cite{zhou2024yolo}, with results shown in Table. \ref{eval-lcs}.
It indicates a clear improvement in the BL metric, which is due to pre-segmenting dysfluency into shorter durations, thus easing the challenge of localizaion. Additionally, Eacc. and Cacc. remain roughly at the same level as the method not using LCS.

\begin{table}[htp!]
    \caption{Evaluation of YOLO-Stutter with or without LCS}
    \label{eval-lcs}
    \centering
    \setlength{\tabcolsep}{5pt} 
    \renewcommand{\arraystretch}{1.2} 
    \resizebox{8.8cm}{!}{
    \small
    \begin{tabular}{l l c c c} 
    \toprule
    Methods & Datasets & Ave. EAcc. ($\%$, $\uparrow$) & \cellcolor{brightturquoise} Ave CAcc.($\%$, $\uparrow$) & \cellcolor{brightturquoise} Ave. BL (ms, $\downarrow$)\\
    \hline
    \hline
    YOLO-Stutter~\cite{zhou2024yolo} & VCTK-Stutter & \textbf{87.57} & 92.79 & 26ms \\
    YOLO-Stutter-LCS & VCTK-Stutter & 83.52 & \textbf{93.28} & \textbf{15ms} \\
    \hline
    YOLO-Stutter~\cite{zhou2024yolo} & VCTK-TTS & \textbf{89.03}  & 94.75 & 31ms \\
    YOLO-Stutter-LCS & VCTK-TTS & 88.66 & \textbf{94.87} & \textbf{23ms} \\
    \bottomrule
    \end{tabular}}
\end{table}

\vspace{-5pt}
\subsection{Token-based Detection}
To assess the performance of the trained Whisper Detector, we evaluated it on the VCTK-Token dataset, as detailed in Table. \ref{token-evaluate}.
From the Table, the low TERs suggests accurate overall transcription. Both EAcc. and CAcc. indicate that model's adeptness at detecting and classifying types of dysfluency, with notably better performance in identifying repetition, pause, and prolongation with relative better performance. Comparison between word and phoneme levels shows that detection at the word level is intuitively easier. 
For TD, the results demonstrate that the model can effectively locate the position of dysfluency. Moreover, localization at the phoneme level proves to be more challenging, as indicated by higher TDs in the table. This increased difficulty is expected, considering that phonemes are smaller units of speech and thus pose greater challenges for precise localization.

\begin{table}[htp!]
    \caption{Evaluation of Whisper Detector on VCTK-Token}
    \label{token-evaluate}
    \centering
    \setlength{\tabcolsep}{5pt} 
    \renewcommand{\arraystretch}{1.2} 
    \resizebox{8.8cm}{!}{
    \small
    \begin{tabular}{l |l c c c c} 
    \toprule
    Levels & Types  & TER ($\%$, $\downarrow$) & EAcc. (\%, $\uparrow$) & CAcc. (\%, $\uparrow$) & TD (e-3, $\downarrow$)\\
    \hline
    \hline
    \multirow{4}{*}{Word} & Repetition & 0.144 & 99.57 & 99.36 & 0.91 \\
    & Deletion & 0.283& 97.98 & 96.85& 1.01\\
    & Insertion& 0.212 & 98.93 & 97.47 & 3.03 \\
    &Pause & 0.195 & 99.04 &98.28 & 7.00\\
    \hline
    \multirow{4}{*}{Phoneme} & Repetition& 0.154 & 98.95& 99.23& 1.87\\
    & Deletion& 1.141 & 98.84& 98.17 & 6.46\\
    & Substitution& 0.335 & 96.11 & 95.75& 7.82 \\
    &Prolongation & 0.326 & 98.89 &98.81 & 2.43 \\
    \bottomrule
    \end{tabular}}
\end{table}

\vspace{-5pt}
\subsection{Unified Benchmark for Time and Token-based Methods}
We conducted a comparative analysis between time-based and token-based methods by evaluating them using both simulated datasets and Aphasia speech data. 
Since VCTK-TTS and VCTK-Token show similar content and dysfluency characteristics (Table. ~\ref{statistics}), we used them as evaluation datasets for each respective method.
The results are detailed in Table. \ref{evaluate-comparison}.
For both simulated datasets and Aphasia speech, the token-based method excels in identifying the existence of dysfluency far more effectively than the time-based method. 
In terms of classification, 
for simulated datasets, the token-based method performs comparably to the time-based method, and it surpasses the time-based method in the types of deletion/missing and replace/substitution.
For Aphasia speech, the token-based method shows significant improvements in addressing missing/deletion and replace/substitution, and it achieves comparable performance in managing repetition.

%% file: conclusion.tex
\section{Conclusion}
In this work, we explored the \textit{token-based} method for dysfluency detection as an alternative to \textit{time-based}~\cite{lian2023unconstrained-udm,lian2024hierarchical,zhou2024yolo} methods. We propose a \textit{token-based} benchmark with decent performance on both simulated data and disordered speech. We unify \textit{time-based} and \textit{token-based} methods with a unified benchmark and show the superiority of the proposed \textit{token-based} methods. However, a more robust evaluation metric is needed to account for dysfluencies occurring before or after text tokens, which current metrics fail to capture. Privacy concerns still necessitate robust de-identification methods.~\cite{gao2021detection}.
Future work will also focus on further scaling efforts at both simulation and model levels. Essentially, we aim to cover many more types of dysfluencies such as filler words, prosody distortion, etc. We also believe this method has the potential to become a framework for a generalized speech rich transcription pipeline. Additionally, we would like to explore fine-grained speech simulation techniques with articulatory features~\cite{lian22bcsnmf, lian2023factor, articulatory_encodec, wu23k_interspeech}.

\section{Acknowledgement}
Thanks for support from UC Noyce Initiative, Society of Hellman Fellows, NIH/NIDCD and the Schwab Innovation fund.

%% file: main.bbl
\begin{thebibliography}{10}
\providecommand{\url}[1]{#1}
\csname url@samestyle\endcsname
\providecommand{\newblock}{\relax}
\providecommand{\bibinfo}[2]{#2}
\providecommand{\BIBentrySTDinterwordspacing}{\spaceskip=0pt\relax}
\providecommand{\BIBentryALTinterwordstretchfactor}{4}
\providecommand{\BIBentryALTinterwordspacing}{\spaceskip=\fontdimen2\font plus
\BIBentryALTinterwordstretchfactor\fontdimen3\font minus \fontdimen4\font\relax}
\providecommand{\BIBforeignlanguage}[2]{{%
\expandafter\ifx\csname l@#1\endcsname\relax
\typeout{** WARNING: IEEEtran.bst: No hyphenation pattern has been}%
\typeout{** loaded for the language `#1'. Using the pattern for}%
\typeout{** the default language instead.}%
\else
\language=\csname l@#1\endcsname
\fi
#2}}
\providecommand{\BIBdecl}{\relax}
\BIBdecl

\bibitem{radford2023whisper}
A.~Radford, J.~W. Kim, T.~Xu, G.~Brockman, C.~McLeavey, and I.~Sutskever, ``Robust speech recognition via large-scale weak supervision,'' in \emph{International conference on machine learning}.\hskip 1em plus 0.5em minus 0.4em\relax PMLR, 2023, pp. 28\,492--28\,518.

\bibitem{zhang2023google-usm}
Y.~Zhang, W.~Han, J.~Qin, Y.~Wang, A.~Bapna, Z.~Chen, N.~Chen, B.~Li, V.~Axelrod, G.~Wang \emph{et~al.}, ``Google usm: Scaling automatic speech recognition beyond 100 languages,'' \emph{arXiv preprint arXiv:2303.01037}, 2023.

\bibitem{pratap2023scaling-speech}
V.~Pratap, A.~Tjandra, B.~Shi, P.~Tomasello, A.~Babu, S.~Kundu, A.~Elkahky, Z.~Ni, A.~Vyas, M.~Fazel-Zarandi \emph{et~al.}, ``Scaling speech technology to 1,000+ languages,'' \emph{arXiv preprint arXiv:2305.13516}, 2023.

\bibitem{lian2023unconstrained-udm}
J.~Lian, C.~Feng, N.~Farooqi, S.~Li, A.~Kashyap, C.~J. Cho, P.~Wu, R.~Netzorg, T.~Li, and G.~K. Anumanchipalli, ``Unconstrained dysfluency modeling for dysfluent speech transcription and detection,'' in \emph{2023 IEEE Automatic Speech Recognition and Understanding Workshop (ASRU)}, 2023, pp. 1--8.

\bibitem{CHIAAI20122157}
\BIBentryALTinterwordspacing
O.~{Chia Ai}, M.~Hariharan, S.~Yaacob, and L.~{Sin Chee}, ``Classification of speech dysfluencies with mfcc and lpcc features,'' \emph{Expert Systems with Applications}, vol.~39, no.~2, pp. 2157--2165, 2012. [Online]. Available: \url{https://www.sciencedirect.com/science/article/pii/S095741741101027X}
\BIBentrySTDinterwordspacing

\bibitem{LPCC}
L.~S. Chee, O.~C. Ai, M.~Hariharan, and S.~Yaacob, ``Automatic detection of prolongations and repetitions using lpcc,'' in \emph{2009 International Conference for Technical Postgraduates (TECHPOS)}, 2009, pp. 1--4.

\bibitem{ESMAILI2016104}
\BIBentryALTinterwordspacing
I.~Esmaili, N.~J. Dabanloo, and M.~Vali, ``Automatic classification of speech dysfluencies in continuous speech based on similarity measures and morphological image processing tools,'' \emph{Biomedical Signal Processing and Control}, vol.~23, pp. 104--114, 2016. [Online]. Available: \url{https://www.sciencedirect.com/science/article/pii/S1746809415001482}
\BIBentrySTDinterwordspacing

\bibitem{10068490}
M.~Jouaiti and K.~Dautenhahn, ``Dysfluency classification in speech using a biological sound perception model,'' in \emph{2022 9th International Conference on Soft Computing \& Machine Intelligence (ISCMI)}, 2022, pp. 173--177.

\bibitem{Detect-lstm}
T.~Kourkounakis, A.~Hajavi, and A.~Etemad, ``Detecting multiple speech disfluencies using a deep residual network with bidirectional long short-term memory,'' in \emph{ICASSP 2020 - 2020 IEEE International Conference on Acoustics, Speech and Signal Processing (ICASSP)}, 2020, pp. 6089--6093.

\bibitem{fluentnet2021}
T.~Kourkounakis and A.~Hajavi, ``Fluentnet: End-to-end detection of stuttered speech disfluencies with deep learning,'' \emph{IEEE/ACM Transactions on Audio, Speech, and Language Processing}, vol.~29, pp. 2986--2999, 2021.

\bibitem{alharbi2020segment-detection3}
S.~Alharbi, M.~Hasan, A.~J. Simons, S.~Brumfitt, and P.~Green, ``Sequence labeling to detect stuttering events in read speech,'' \emph{Computer Speech \& Language}, vol.~62, p. 101052, 2020.

\bibitem{segment-detection4}
M.~Jouaiti and K.~Dautenhahn, ``Dysfluency classification in stuttered speech using deep learning for real-time applications,'' in \emph{ICASSP 2022 - 2022 IEEE International Conference on Acoustics, Speech and Signal Processing (ICASSP)}, 2022, pp. 6482--6486.

\bibitem{DBLP:conf/interspeech/BayerlWNR22}
\BIBentryALTinterwordspacing
S.~P. Bayerl, D.~Wagner, E.~N{\"{o}}th, and K.~Riedhammer, ``Detecting dysfluencies in stuttering therapy using wav2vec 2.0,'' in \emph{Interspeech 2022, 23rd Annual Conference of the International Speech Communication Association, Incheon, Korea, 18-22 September 2022}.\hskip 1em plus 0.5em minus 0.4em\relax {ISCA}, 2022, pp. 2868--2872. [Online]. Available: \url{https://doi.org/10.21437/Interspeech.2022-10908}
\BIBentrySTDinterwordspacing

\bibitem{howell1995automatic}
P.~Howell and S.~Sackin, ``Automatic recognition of repetitions and prolongations in stuttered speech,'' in \emph{Proceedings of the first World Congress on fluency disorders}, vol.~2.\hskip 1em plus 0.5em minus 0.4em\relax University Press Nijmegen Nijmegen, The Netherlands, 1995, pp. 372--374.

\bibitem{10094692}
P.~Mohapatra, B.~Islam, M.~T. Islam, R.~Jiao, and Q.~Zhu, ``Efficient stuttering event detection using siamese networks,'' in \emph{ICASSP 2023 - 2023 IEEE International Conference on Acoustics, Speech and Signal Processing (ICASSP)}, 2023, pp. 1--5.

\bibitem{lian2024hierarchical}
J.~Lian and G.~Anumanchipalli, ``Towards hierarchical spoken language dysfluency modeling,'' \emph{Proceedings of the 18th Conference of the European Chapter of the Association for Computational Linguistics}, 2024.

\bibitem{zhou2024yolo}
X.~Zhou, A.~Kashyap, S.~Li, A.~Sharma, B.~Morin, D.~Baquirin, J.~Vonk, Z.~Ezzes, Z.~Miller, M.~L.~G. Tempini \emph{et~al.}, ``Yolo-stutter: End-to-end region-wise speech dysfluency detection,'' \emph{Interspeech}, 2024.

\bibitem{redmon2016look}
J.~Redmon, S.~Divvala, R.~Girshick, and A.~Farhadi, ``You only look once: Unified, real-time object detection,'' 2016.

\bibitem{kim2021conditional-vits}
J.~Kim, J.~Kong, and J.~Son, ``Conditional variational autoencoder with adversarial learning for end-to-end text-to-speech,'' in \emph{International Conference on Machine Learning}.\hskip 1em plus 0.5em minus 0.4em\relax PMLR, 2021, pp. 5530--5540.

\bibitem{radford2022whisper}
A.~Radford, J.~W. Kim, T.~Xu, G.~Brockman, C.~McLeavey, and I.~Sutskever, ``Robust speech recognition via large-scale weak supervision,'' in \emph{International Conference on Machine Learning}.\hskip 1em plus 0.5em minus 0.4em\relax PMLR, 2023, pp. 28\,492--28\,518.

\bibitem{wiki-lcs}
``Longest common subsequence,'' \url{https://en.wikipedia.org/wiki/Longest \\ \_common\_subsequence}.

\bibitem{cmu-phoneme-dict}
``Cmu phoneme dictionary,'' \url{http://www.speech.cs.cmu.edu/cgi-bin/cmudict}.

\bibitem{yamagishi2019cstr-vctk}
J.~Yamagishi, C.~Veaux, K.~MacDonald \emph{et~al.}, ``Cstr vctk corpus: English multi-speaker corpus for cstr voice cloning toolkit (version 0.92),'' \emph{University of Edinburgh. The Centre for Speech Technology Research (CSTR)}, 2019.

\bibitem{gao2021detection}
Y.~Gao, J.~Lian, B.~Raj, and R.~Singh, ``Detection and evaluation of human and machine generated speech in spoofing attacks on automatic speaker verification systems,'' in \emph{2021 IEEE Spoken Language Technology Workshop (SLT)}.\hskip 1em plus 0.5em minus 0.4em\relax IEEE, 2021, pp. 544--551.

\bibitem{lian22bcsnmf}
J.~Lian, A.~W. Black, L.~Goldstein, and G.~K. Anumanchipalli, ``{Deep Neural Convolutive Matrix Factorization for Articulatory Representation Decomposition},'' in \emph{Proc. Interspeech 2022}, 2022, pp. 4686--4690.

\bibitem{lian2023factor}
J.~Lian, A.~W. Black, Y.~Lu, L.~Goldstein, S.~Watanabe, and G.~K. Anumanchipalli, ``Articulatory representation learning via joint factor analysis and neural matrix factorization,'' in \emph{ICASSP 2023-2023 IEEE International Conference on Acoustics, Speech and Signal Processing (ICASSP)}.\hskip 1em plus 0.5em minus 0.4em\relax IEEE, 2023, pp. 1--5.

\bibitem{articulatory_encodec}
C.~J. Cho, P.~Wu, T.~S. Prabhune, D.~Agarwal, and G.~K. Anumanchipalli, ``Articulatory encodec: Vocal tract kinematics as a codec for speech,'' \emph{arXiv preprint arXiv:2406.12998}, 2024.

\bibitem{wu23k_interspeech}
P.~Wu, T.~Li, Y.~Lu, Y.~Zhang, J.~Lian, A.~W. Black, L.~Goldstein, S.~Watanabe, and G.~K. Anumanchipalli, ``{Deep Speech Synthesis from MRI-Based Articulatory Representations},'' in \emph{Proc. INTERSPEECH 2023}, 2023, pp. 5132--5136.

\end{thebibliography}
